\documentclass[%
 reprint,
 superscriptaddress,
 longbibliography,
 amsmath,amssymb,
 aps,
prb,
]{revtex4-2}

\usepackage{graphicx}
\usepackage{dcolumn}
\usepackage{bm}

\usepackage{xcolor}
\usepackage{soul}
\usepackage{physics}
\usepackage{bbold}

\begin{document}

\preprint{APS/123-QED}


\title{Quantum sensing of electric field distributions of liquid electrolytes with NV-centers in nanodiamonds}
\author{M. Hollendonner}
\affiliation{Friedrich-Alexander-University Erlangen-Nuremberg, 91058 Erlangen, Germany}
\affiliation{Max Planck Institute for the Science of Light, 91058 Erlangen, Germany}

\author{S. Sharma}
\affiliation{Max Planck Institute for the Science of Light, 91058 Erlangen, Germany}

\author{D. B. R. Dasari}
\affiliation{3rd Institute of Physics, IQST, and Research Center SCoPE, University of Stuttgart, 70569 Stuttgart, Germany}

\author{A. Finkler}
\affiliation{Department of Chemical and Biological Physics, Weizmann Institute of Science, Rehovot 7610001, Israel}

\author{S. V. Kusminskiy}
\affiliation{Max Planck Institute for the Science of Light, 91058 Erlangen, Germany}
\affiliation{Institute for Theoretical Solid State Physics, RWTH Aachen University, 52074 Aachen, Germany}

\author{R. Nagy}
\email{roland.nagy@fau.de}
\affiliation{Friedrich-Alexander-University Erlangen-Nuremberg, 91058 Erlangen, Germany}

\date{\today}

\begin{abstract}
To use batteries as large-scale energy storage systems it is necessary to measure and understand their degradation \textit{in-situ} and \textit{in-operando}. As a battery's degradation is often the result of molecular processes inside the electrolyte, a sensing platform which allows to measure the ions with a high spatial resolution is needed. Primary candidates for such a platform are NV-centers in diamonds. We propose to use a single NV-center to deduce the electric field distribution generated by the ions inside the electrolyte through microwave pulse sequences. We show that the electric field can be reconstructed with great accuracy by using a protocol which includes different variations of the Free Induction Decay to obtain the mean electric field components and a modified Hahn-echo pulse sequence to measure the electric field's standard deviation $\sigma_E$. From a semi-analytical ansatz we find that for a lithium ion battery there is a direct relationship between $\sigma_E$ and the ionic concentration. Our results show that it is therefore possible to use NV-centers as sensors to measure both the electric field distribution and the local ionic concentration inside electrolytes.
\end{abstract}

\maketitle


\section{\label{sec:level1}Introduction}
Rechargeable batteries play an important role for our society and are a key ingredient for the transition towards renewable energy sources \cite{dioufPotentialLithiumionBatteries2015a, dilecceLithiumionBatteriesSustainable2017, jaiswalLithiumionBatteryBased2017}. As the production of batteries is accompanied with a considerable use of resources, recyclable \cite{harperRecyclingLithiumionBatteries2019} batteries with a long lifetime are needed. The latter is limited by degradation mechanisms, such as the formation of solid-electrolyte interfaces \cite{mengDesigningBetterElectrolytes2022} or lithium-plating \cite{medaSolidElectrolyteInterphase2022} which can reduce the battery's capacity with increasing cell age \cite{edgeLithiumIonBattery2021}. As these processes happen on a molecular level within nanometer scales \cite{mengDesigningBetterElectrolytes2022}, a sensor which is capable of monitoring the ionic concentration \textit{in-situ} and \textit{in-operando} with high spatial and temporal resolutions is needed. Even though MRI allows to reconstruct transport properties \cite{klettQuantifyingMassTransport2012,krachkovskiyVisualizationSteadyStateIonic2016} of a battery, tools which allow to perform measurements inside the electrolyte are still absent \cite{mengDesigningBetterElectrolytes2022}. 

It has been demonstrated that nitrogen-vacancy (NV) centers in diamond (see Fig.\,\ref{fig:setting}(b)) are high-resolution quantum sensors, which can detect oscillating or fluctuating \cite{hallSensingFluctuatingNanoscale2009,steinertMagneticSpinImaging2013,luanDecoherenceImagingSpin2015,agarwalMagneticNoiseSpectroscopy2017} magnetic fields with nano- \cite{balasubramanianUltralongSpinCoherence2009,webbNanoteslaSensitivityMagnetic2019} and even subpico-Tesla \cite{wolfSubpicoteslaDiamondMagnetometry2015} sensitivities. Besides this, NV-centers have great ability for the detection of electric fields. They can not only detect DC \cite{doldeElectricfieldSensingUsing2011,bianNanoscaleElectricfieldImaging2021} or AC \cite{michlRobustAccurateElectric2019a} electric fields with remarkable precision, but are additionally capable of detecting single fundamental charges \cite{doldeNanoscaleDetectionSingle2014} even within the diamond lattice \cite{mittigaImagingLocalCharge2018}. This electric field sensitivity was used by Ref. \cite{dinaniSensingElectrochemicalSignals2021} to show that, based on theoretical considerations, bulk NV-centers can work as electrochemical sensors if they are in contact with an electrolyte solution.

Here we show that nanodiamonds equipped with single NV-centers can act as \textit{in-situ} electric field sensors inside liquid electrolytes (Fig.\,\ref{fig:setting}(a)). By exploiting how transverse and axial electric fields act on the NV-center's ground state spin states, we find variations of the free-induction decay (FID) pulse sequence, which allow to measure the mean electric field components. Further, we show that it is possible to use variants of the Hahn-echo pulse sequence to additionally obtain the electric field's standard deviation $\sigma_E$. From a semi-analytical ansatz we demonstrate exemplarily for a lithium ion battery (LIB) that there is a direct relationship between the electric field's standard deviation and the local ionic concentration. A nanodiamond with a single NV-center can therefore work as a sensor which allows to simultaneously reconstruct the electric field distribution and to measure the ionic concentration with $\mathrm{nm}$ spatial resolution.

\section{\label{sec:el_field_distribution_LIB}Electric field distribution in liquid electrolytes}

\begin{figure*}
    \includegraphics[width=0.95\textwidth]{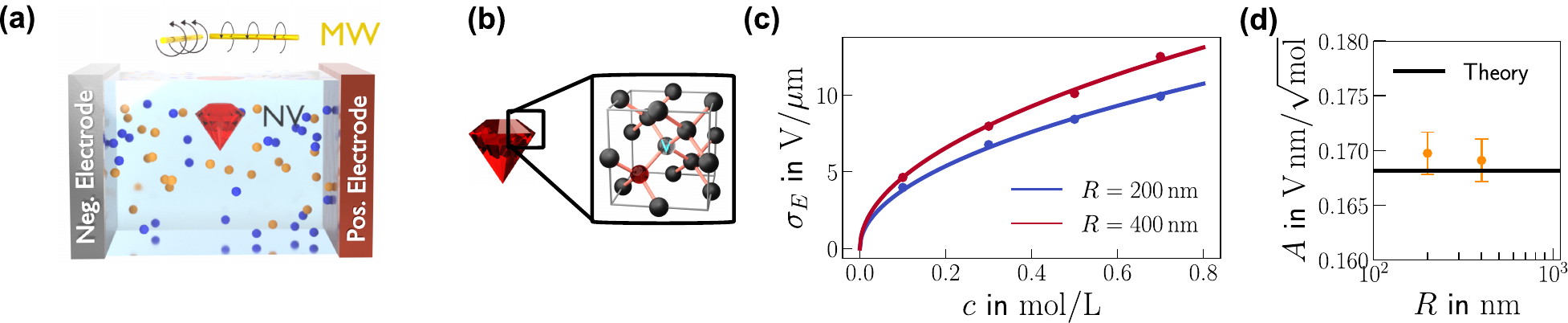}
    \caption{\label{fig:setting}(a) Experimental setting. A nanodiamond which is dissolved in the liquid electrolyte of the battery is surrounded by positive (orange) and negative (blue) ions. Two perpendicular aligned gold wires allow to generate polarized microwave drives. (b) To work as a quantum sensor, the nanodiamond contains a vacancy (V) next to a nitrogen atom (red). (c) Standard deviation of $E_z$, calculated from 500 repeated sets of randomly placed ions of concentration $c$ around the nanodiamond ($r_{\mathrm{ND}}=100\,\mathrm{nm}$) and inside a sphere of radius $R$. The relative permittivities are $\epsilon_{\mathrm{ND}}=5.8$ \cite{dinaniSensingElectrochemicalSignals2021} and $\epsilon_e=17.5$ \cite{liuStructuresDynamicProperties2019}. Solid lines are fits following Eq.\,(\ref{eq:sigma_Ez}) with $A$ as a fit parameter. (d) Fit parameters $A$ obtained from (c), compared to the theory value.}
\end{figure*}

Before introducing measurements of the electric field distribution by the NV-center, we would like to develop an analytic expression of the electric field induced inside the nanodiamond by the positive and negative ions of the electrolyte.

The potential $\Phi$ at position $\mathbf{r}$ inside the nanodiamond due to a single charge $q$ at position $\mathbf{b}$, is described by Poisson's equation
\begin{equation}
    \nabla^2 \Phi \left(\mathbf{r}\right) = -\frac{\rho \left(\mathbf{r}\right)}{\epsilon}\,. \label{eq:Poisson}
\end{equation}
Here $\epsilon = \epsilon_0 \epsilon_i$ with $i=e,\,\mathrm{ND}$, are the permittivities of, respectively, the electrolyte and the nanodiamond in terms of the vacuum permittivity $\epsilon_0$ and $\rho$ is the charge density induced by $q$. The solution inside the nanodiamond, $\Phi_\mathrm{ND}$ (see Methods for the detailed derivation), allows to obtain the electric field at the center of the nanodiamond, which is
\begin{equation}
    \mathbf{E}^{ND}=\frac{q}{4\pi\epsilon_{0}}\frac{3}{2\epsilon_{e}+\epsilon_{\mathrm{ND}}}\frac{\mathbf{b}}{b^{3}}\,.\label{eq:E_ND}
\end{equation}
By considering the positions of ions of a molar concentration $c$ to be normally distributed within a sphere of radius $R$ around a nanodiamond (radius $r_{\mathrm{ND}}$), the standard deviation of the electric field distribution at the center of the nanodiamond is
\begin{align}
    \sigma_{E_{z}} &= A\sqrt{c\left(\frac{1}{r_{\mathrm{ND}}}-\frac{1}{R}\right)} \nonumber\\
    A &= \frac{|q|}{\epsilon_{0}\left(2\epsilon_{e}+\epsilon_{\mathrm{ND}}\right)}\sqrt{\frac{3N_{A}}{4\pi}}\,. \label{eq:sigma_Ez}
\end{align}
To validate Eq.\,(\ref{eq:sigma_Ez}), we simulated the standard deviation of 500 sets of uniformly and randomly placed ions for different molar ionic concentrations (see Fig.\,\ref{fig:setting}(c)). As it is the most widely used electrolyte of LIBs \cite{marcinekElectrolytesLiionTransport2015}, we chose $\mathrm{LiPF_6^-}$ with $\epsilon_e=17.5$ \cite{liuStructuresDynamicProperties2019}. The total electric field was calculated as the linear sum of Eq.\,(\ref{eq:E_ND}) for all randomly placed ions around a $200\,$nm spherical nanodiamond \cite{woodLongSpinCoherence2022}. As it can be seen from Fig.\,\ref{fig:setting}(d), the expected $A$ value is in fair agreement with the simulations. From Eq.\,(\ref{eq:sigma_Ez}) it can be calculated that for $R=500\,\mathrm{nm}$, the fluctuations will increase only by 3$\%$, compared to $\sigma_E\left(R=400\,\mathrm{nm}\right)$. As $\sigma_E$ therefore saturates for $R\gtrsim 500\,\mathrm{nm}$, this implies that electric field fluctuations only affect the nanodiamond within sub-micrometer range and the system is limited by the confocal volume of the experimental setup, which typically is $\sim 1\,\mathrm{\mu m}^3$ \cite{misonouConstructionOperationTabletop2020, maertzVectorMagneticField2010}.

\section{\label{sec:sensing_el_fields_in_LIB}Sensing of static electric fields inside electrolytes}

\begin{figure*}
    \includegraphics[width=0.93\textwidth]{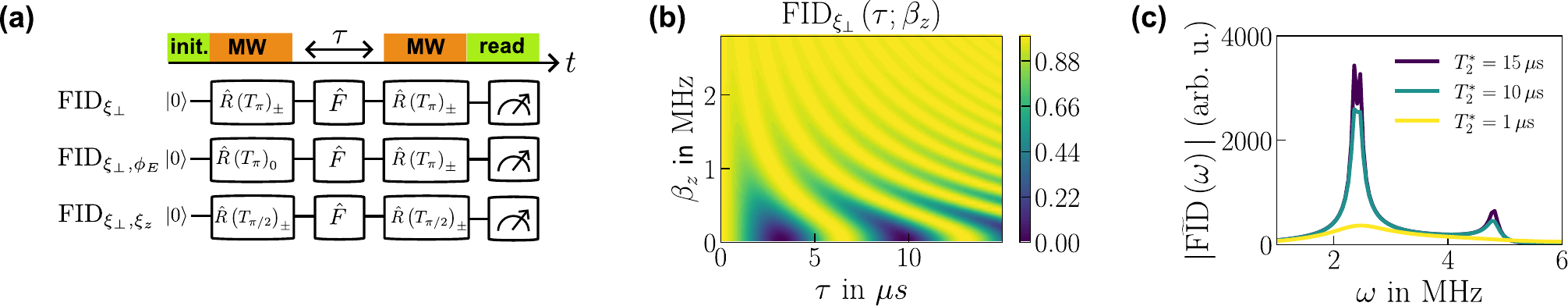}
    \caption{\label{fig:FID_wide_figure}(a) FID-variations to extract $\xi_\perp$, $\phi_E$ and $\xi_z$ through subsequent pulse sequences. Here $T_\pi$ ($T_{\pi/2}$) is the duration of the microwave pulse such that a $\pi$-pulse ($\pi/2$-pulse) is performed. Subscripts $\pm$ denote circularly polarized drives which cause oscillations between $\ket{0}$ and either $\ket{1}$ or $\ket{-1}$. Subscript $0$ denotes linear polarization of the drive and the free evolution is described through $\hat{F}$. (b) $\mathrm{FID}_{\xi_\perp}$ for different magnetic fields up to $\beta_z=2.7\,\mathrm{MHz}$, corresponding to $B_z=1\,\mathrm{G}$. For $\beta_z=0$ the signal has the highest contrast with the lowest frequency of oscillation. (c) Fourier transform of $\mathrm{FID}_{\xi_\perp,\xi_z}$ with $\Omega=10\,\mathrm{MHz}$ and $E_{x,y,z}=10\,\mathrm{V/\mu m}$. Only for $T_2^* > 10\,\mathrm{\mu s}$ the peaks at $\xi_\perp \pm \xi_z = 2.4 \pm 0.04\,\mathrm{MHz}$ and $2\xi_\perp$ can be resolved.}
\end{figure*}

An electric field $\mathbf{E}$ can in cylindrical coordinates be expressed by its axial component $E_z$, its transverse projection $E_\perp = \sqrt{E_x^2 + E_y^2}$ and an angle $\phi_E$, which defines the projections onto the $x$ and $y$ axis as $E_x = E_\perp \cos \phi_E$ and $E_y = E_\perp \sin \phi_E$. The total Hamiltonian which describes the NV-center in presence of electric and axial magnetic fields will in the following be denoted as $\hat{\mathcal{H}}_0$. By taking into account that the NV-center can be driven by two perpendicular microwave wires (see Fig.\,\ref{fig:setting}(a)) with amplitude $\Omega$, frequency $\omega_d$ and a phase $\phi$ between each other, the total ground state Hamiltonian in a frame rotating  with $\omega_d$ is $\hat{\mathcal{H}}= \hat{\mathcal{H}}_0 + \hat{\mathcal{H}}_d$ (see Methods), where
\begin{align}
     & \nonumber\\
    \hat{\mathcal{H}}_0 &= \left(\Delta + \xi_z\right)\hat{S}_z^2 + \beta_z \hat{S}_z -\frac{\xi_\perp}{2}\left(\hat{S}_+^2 e^{i\phi_E} + h.c.\right) \nonumber\\
    \hat{\mathcal{H}}_d &= \frac{\Omega}{\sqrt{2}}\left(\epsilon_-\sigma_{0,-1}+\epsilon_+\sigma_{0,+1}^\dagger+h.c.\right)\,.\label{eq:H_rwa}
\end{align}
Here $\Delta=D-\omega_{d}$ is the detuning between the zero-field splitting, $D=2.87\,\mathrm{GHz}$ \cite{loubserElectronSpinResonance1978}, and the microwave drive frequency. $S_i$, $i=x,y,z$, are the spin-1 operators, which can be used to define ladder operators $S_\pm = S_x \pm i S_y$. $\sigma_{0,\pm 1}=\ket{0}\bra{\pm 1}$ are operators which describe transitions between $\ket{0}$ and, respectively, $\ket{\pm 1}$. Frequency contributions generated by electric and axial magnetic fields are considered through $\xi_z = d_\parallel E_z$ and $\xi_\perp = d_\perp E_\perp$ ($d_\parallel = 0.35\,\mathrm{Hz\, cm/V}$, $d_\perp = 17\,\mathrm{Hz\, cm/V}$ \cite{vanoortElectricfieldinducedModulationSpin1990}) and $\beta_z = \gamma_e B_z$ ($\gamma_e=28\,\mathrm{GHz/T}$ \cite{abeTutorialMagneticResonance2018}). 

The phase factors $\epsilon_\pm = \left(1-ie^{\mp i\phi}\right)/2$ which enter into Eq.\,(\ref{eq:H_rwa}), allow to describe the transitions which are caused by circularly ($\phi=\pm \pi/2$) or linearly ($\phi=0$) polarized microwave drives \cite{londonStrongDrivingSingle2014}. The time-evolution operators of $\hat{\mathcal{H}}_d$, $\hat{R}\left(t\right)=e^{-i\hat{\mathcal{H}}_dt}$ (see Methods), show that one can induce Rabi oscillations between $\ket{0}$ and $\ket{1}$ for right circularly polarized drives and $\ket{0}\leftrightarrow \ket{-1}$ for left circular polarizations. Linearly polarized drives allow to drive transitions between $\ket{0}$ and both $\ket{\pm 1}$.

In absence of microwave drives, the $\ket{\pm 1}$ states are symmetrically mixed by $\xi_\perp$ and axial electric fields effectively shift $\ket{0}$ from $\ket{\pm 1}$, which can be seen from $\hat{F}\left(\tau\right) = e^{-i\hat{\mathcal{H}}_0 \tau}$ (see Methods). As axial and transverse electric fields thus act differently on the $\ket{m_s = 0,\pm 1}$ states of the NV-center, one can derive variations of the Free Induction Decay (FID), which allow to extract these electric field components.

\subsection{\label{sec:Sensing_E-field_components}Measurement of electric field components}

The FID consists of two microwave pulses separated by a free evolution period $\tau$. Electric field contributions $\xi_\perp$, $\phi_E$ and $\xi_z$ can be sensed through FID-variations, as shown in Fig.\,\ref{fig:FID_wide_figure}(a). The NV-center can be initialized into its $\ket{0}$ state via excitation with green laser light, followed by intersystem-crossing \cite{dohertyNitrogenvacancyColourCentre2013}. This state can then be driven to $-i\ket{1}$ through a right-polarized $\pi$-pulse, denoted as $\hat{R}\left(T_\pi\right)_+$, and will be influenced by both axial magnetic as well as transverse electric fields. The latter induce mixing with $\ket{-1}$. By using a microwave $\pi$-pulse with the same polarization as the initial one, the transferred population from $\ket{1}$ to $\ket{-1}$ can be obtained from the FID-signal
\begin{align}
    \mathrm{FID}_{\xi_\perp}\left(\tau\right) &= |\bra{0}\hat{R}\left(T_\pi\right)_+\hat{F}\left(\tau\right)\hat{R}\left(T_\pi\right)_+\ket{0}|^2 \nonumber\\
    &= \cos^2\left(\tau \sqrt{\beta_z^2 + \xi_\perp^2}\right) \nonumber\\
    &+ \frac{\beta_z^2}{\beta_z^2 + \xi_\perp^2}\sin^2\left(\tau \sqrt{\beta_z^2 + \xi_\perp^2}\right)\,,\label{eq:FID_xi_perp}
\end{align}
which is a measure of the population which has been transferred from $\ket{1}$ to $\ket{-1}$. In Fig.\,\ref{fig:FID_wide_figure}(b) one can see this FID-signal as a function of the free evolution time $\tau$ for $\beta_z$ values up to $2.8\,\mathrm{MHz}$, which corresponds to $B_z=1\,\mathrm{G}$. Besides having a decreased contrast for $\beta_z \neq 0$, the frequency $\sqrt{\beta_z^2 + \xi_\perp^2}$ of the FID-oscillations depends on both axial magnetic and transverse electric fields. It is therefore strongly recommended to perform the measurements in a magnetically shielded environment, for example by a $\mu$-metal as in Ref. \cite{zhaoAtomicscaleMagnetometryDistant2011}. In the following it will be assumed that all measurement are performed without any magnetic field being present.

The transverse electric field components are uniquely defined through $\phi_E$, as $\xi_x = \xi_\perp \cos \phi_E$ and $\xi_y = \xi_\perp \sin \phi_E$. A superposition state $-e^{i\pi/4}\left(\ket{1}+\ket{-1}\right)/\sqrt{2}$ generated through a linearly polarized $\pi$-pulse (considered via $\hat{R}\left(T_\pi\right)_0$, see Methods) will additionally to $\xi_\perp$ also be affected by $\phi_E$ as this phase differs in its sign for $\ket{1}$ and $\ket{-1}$ (see Methods). If either $\ket{1}$ or $\ket{-1}$ is projected to $\ket{0}$ through the final microwave pulse, one obtains an FID-signal, which both depends on $\xi_\perp$ and $\phi_E$,
\begin{equation}
    \mathrm{FID}_{\phi_E, \xi_\perp}\left(\tau\right) = \frac{1}{2}\left(1-\sin\left(2\tau\xi_\perp\right)\sin \phi_E\right)\,.\label{eq:FID_phiE,xi_perp}
\end{equation}
One can obtain $\phi_E$ as the relative fraction between the value of the FID-signal at $\tau=0$ and its first maxima at $2\tau\xi_\perp = \pi/2$,
\begin{equation}
    \frac{\mathrm{FID}_{\phi_E, \xi_\perp}\left(\tau=\frac{\pi}{2}\frac{1}{2\xi_\perp}\right)}{\mathrm{FID}_{\phi_E, \xi_\perp}\left(\tau=0\right)} = 1-\sin \phi_E\,.
\end{equation}
By using $\mathrm{FID}_{\xi_\perp}$ and $\mathrm{FID}_{\xi_\perp,\phi_E}$, it is therefore possible to not only determine the electric field's transverse component, but also to obtain the projection onto the $x$ and $y$ axes, which are determined through $\phi_E$.

Axial electric field contributions $\xi_z$ cause a Stark shift between $\ket{0}$ and $\ket{\pm 1}$. A superposition state $\left(\ket{0} -i \ket{-1}\right)/\sqrt{2}$ generated by a circularly polarized $\pi/2$-pulse (see Fig.\,\ref{fig:FID_wide_figure}(a)) will therefore be affected both by $\xi_z$ and $\xi_\perp$. If the final microwave $\pi/2$-pulse has the same polarization as the initial one, an FID-signal is obtained which depends both on $\xi_\perp$ and $\xi_z$, 
\begin{equation}
    \mathrm{FID}_{\xi_z,\xi_\perp}\left(\tau\right) = \frac{1}{4}\left(1-2\cos\left(\tau \xi_\perp\right)\cos\left(\tau \xi_z\right) + \cos^2\left(\tau \xi_\perp\right)\right)\,,\label{eq:FID_xi_perp,xi_z}
\end{equation}
if the NV-center was driven with $\omega_d = D$. The Fourier transform of Eq.\,(\ref{eq:FID_xi_perp,xi_z}) (see Methods),
\begin{align}
    \widetilde{\mathrm{FID}}\left(\omega > 0\right) &= \frac{\pi}{4}\Bigl(\frac{1}{2}\delta\left(2\xi_\perp - \omega\right) \nonumber\\ &-\delta\left(\xi_\perp + \xi_z - \omega\right) -\delta\left(\xi_\perp - \xi_z - \omega\right)\Bigr)\,,\label{eq:FT_FID_xip_phiE}
\end{align}
shows, that $\xi_z$ can be measured if it is possible to spectrally resolve $\xi_\perp \pm \xi_z$. To study this, we numerically \cite{johanssonQuTiPOpensourcePython2012,johanssonQuTiPPythonFramework2013} simulated $\mathrm{FID}_{\xi_z,\xi_\perp}$ and included dephasing at rates $1/T_2^*$ through a Lindblad operator $\sqrt{1/T_2^*} S_z$ for $T_2^*$ in the range up to $15\,\mathrm{\mu s}$ (see Fig.\,\ref{fig:FID_wide_figure}(c)). One can resolve $\xi_\perp \pm \xi_z$ for nanodiamonds with $T_2^* > 10\,\mathrm{\mu s}$, which is higher than the value of typical nanodiamonds \cite{knowlesObservingBulkDiamond2014}. For a nanodiamond with $T_2^* \approx 15\,\mathrm{\mu s}$ it would be possible to distinguish between $\xi_\perp$ and $\xi_z$ and therefore to determine the projection of the electric field onto the symmetry axis of the NV-center.

\section{\label{sec:fid_fluctuations}Influence of fluctuating electric fields}

\begin{figure}[h]
    \includegraphics[width=0.45\textwidth]{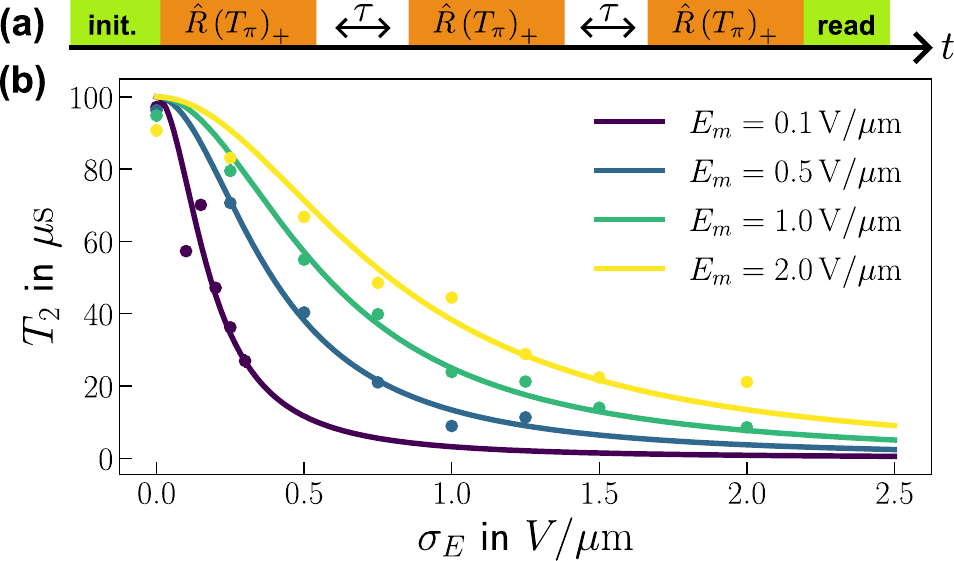}
    \caption{\label{fig:hahn-echo}(a) Hahn-echo pulse sequence used to simulate Eq.\,(\ref{eq:Average_Hahn-echo}). (b) Total $T_2$ for numerically \cite{johanssonQuTiPOpensourcePython2012,johanssonQuTiPPythonFramework2013} simulated Hahn-Echoes with $T_{2,int}=100\,\mathrm{\mu s}$, with the electric field components sampled from a normal distribution with mean $E_m$ and standard deviation $\sigma_E$. For the simulations a drive of $\Omega=10\,$MHz was used. Solid lines are fits of $\alpha E_m/\sigma_E^2$. Every trajectory was obtained from 1000 individual simulations. Error bars of one standard deviation are smaller than the data points.}
\end{figure}

It can be assumed that the ions surrounding the nanodiamond will not stay static for the timescales in which measurements are performed but will be subject to, for instance, drift and diffusion. These fluctuations will affect the electric field inside the nanodiamond. Due to the limited $T_2^*$ of nanodiamonds, the FID pulse sequences as introduced before will be mainly suitable for the measurement of the average electric fields (see Methods). 
The coherence time of a nanodiamond can be significantly prolonged if instead of an FID, a Hahn-Echo pulse sequence is used \cite{woodLongSpinCoherence2022}. As it is shown in Fig.\,\ref{fig:hahn-echo}(a), we propose a modified version of the Hahn-Echo, where after the first free evolution interval, a $\pi$-pulse with right-circular polarization is performed, before the spin is allowed to precess freely during a second free evolution interval $\tau$. Before being read out, a right-circularly polarized $\pi$-pulse is applied, which leads to a signal $\mathrm{Hahn}\left(\tau\right)= \left(1-\cos\left(2\tau \xi_\perp\right)\right)^2/4$.
Simulations of this Hahn-Echo variation show that the averages (see Methods for an example) can be fit by
\begin{equation}
    \left\langle \mathrm{Hahn}\left(\tau\right)\right\rangle = \frac{1}{4}\left[1-\cos\left(2\tau\xi_{\perp}\right)e^{-\tau/T_{2}}\right]^{2}\,.\label{eq:Average_Hahn-echo}
\end{equation}
Here $T_{2}$ is the sum of the intrinsic spin coherence time $T_{2,int.}=100\,\mathrm{\mu s}$ \cite{woodLongSpinCoherence2022} and a contribution due to the fluctuating electric fields,
\begin{align}
    \frac{1}{T_{2}} &= \frac{1}{T_{2,int.}} + \frac{1}{T_{2,E}}\,.\label{eq:T2_tot}
\end{align}
In Fig.\,\ref{fig:hahn-echo}(b), one can see $T_2$ as a function of the electric field's standard deviation $\sigma_E$, where solid lines are $T_{2,E} = \alpha E_m/\sigma_E^2$ in terms of a fit parameters $\alpha$. The total spin coherence time is therefore strongly affected by $\sigma_E$ and the mean electric field value $E_m$. If the mean transverse electric field has been sensed by the FID sequence as shown in Eq.\,(\ref{eq:FID_xi_perp}), it is therefore possible to derive the electric field's standard deviation, which together with $\xi_\perp$, $\phi_E$ and $\xi_z$ defines the electric field distribution. As there is a direct relationship between $\sigma_E$ and the local ionic concentration (see Fig.\,\ref{fig:setting}(c)), the proposed Hahn-echo pulse sequence additionally allows to use the NV-center inside the nanodiamond as a local concentration sensor.

\section{\label{sec:conclusion_and_outlook}Conclusion and Outlook}
 In conclusion we have shown here a full reconstruction of the mean electric field generated in a liquid electrolyte, through the spin control of a quantum sensor immersed in the electrolyte. We have found exact expressions correlating the electric field components with the free-induction decay of the sensor spin, and the dependence of the variance on the spin-echo measurements. Together we were able to deduce the electric field distribution and also measure the local ionic concentration, a key parameter in characterizing the performance of the liquid electrolyte for battery applications. We envisage that with improved modeling of the electric field distribution in liquid electrolytes and using better quantum control methods, for example using correlation spectroscopy \cite{laraouiHighresolutionCorrelationSpectroscopy2013}, we could enhance the sensitivity of the sensor to the local electric-field environment, allowing for an \textit{in-situ} monitoring of the battery using the liquid electrolyte.

\begin{acknowledgments}
R. N. would like to acknowledge financial support by the Federal Ministry of Education and Research (BMBF) project QMNDQCNet and DFG (Project No. 507241320 and 46256793). S. V. K. and D. D. would like to acknowledge the funding support from BMBF (Grant No. 16KIS1590K). A. F. is the incumbent of the Elaine Blond Career Development Chair and acknowledges support from Israel Science Foundation (ISF grants 963/19 and 418/20) as well as the Abramson Family Center for Young Scientists and the Willner Family Leadership Institute for the Weizmann Institute of Science.
\end{acknowledgments}

\newpage

\bibliographystyle{ieeetr}
\bibliography{NV-Centers.bib, Software.bib, Batteries.bib, Books.bib}
\clearpage
\newpage
\widetext

\begin{center}
    \textbf{\large Quantum sensing of electric field distributions of liquid electrolytes with NV-centers in nanodiamonds - Supplementary Information}
\end{center}

\setcounter{equation}{0}
\setcounter{figure}{0}
\setcounter{table}{0}
\setcounter{page}{1}
\setcounter{section}{0}
\makeatletter
\renewcommand{\theequation}{S\arabic{equation}}
\renewcommand{\thefigure}{S\arabic{figure}}

\section{Electric field at center of nanodiamond}\label{App:E_ND}
In the following we would like to deduce the electric field of a single point charge $q$ at a distance $\mathbf{b}$ from the origin of the nanodiamond with radius $r_{\mathrm{ND}}$ by following Ref.\,[S1]. Poisson's equation describes the electrostatic potential $\Phi$,
\begin{equation}
    \nabla^2 \Phi \left( \mathbf{r} \right) = -\frac{\rho\left(\mathbf{r}\right)}{\epsilon}\,,
\end{equation}
where $\epsilon = \epsilon_0 \epsilon_i$, $i=e,\,\mathrm{ND}$ is the permittivity of, respectively, the electrolyte and the nanodiamond in terms of the vacuum permittivity $\epsilon_0$. By exploiting azimuthal symmetry of the problem, the above expression reduces to Laplace's equation for $\mathbf{r}\neq \mathbf{b}$, which in spherical coordinates with $|\mathbf{r}|=r$ and $\theta$ the angle spanned by $\mathbf{r}$ and $\mathbf{b}$ is
\begin{equation}
    \nabla^{2}\Phi\left(r,\theta\right)=\frac{1}{r^{2}}\frac{\partial}{\partial r}\left(r^{2}\frac{\partial\Phi}{\partial r}\right)+\frac{1}{r^{2}\sin\theta}\frac{\partial}{\partial\theta}\left(\sin\theta\frac{\partial\Phi}{\partial\theta}\right)=0\,.
\end{equation}
The general solution of this partial differential equation can be expressed in terms of the associated Legendre polynomials $P_l$ of order $l$ and in terms of two constants $A_l$ and $C_l$ as [S1,\,S2]
\begin{equation}
    \Phi\left(r,\theta\right)=\sum_{l=0}^{\infty}\left(A_{l}r^{l}+C_{l}\frac{1}{r^{l+1}}\right)P_{l}\left(\cos\theta\right)\,.\label{eq:Phi_ansatz}
\end{equation}
As the potential inside the nanodiamond must be finite at $r=0$, $C_l$ needs to vanish and one therefore has
\begin{equation}
    \Phi^{\mathrm{ND}}\left(r,\theta\right) = \sum_{l=0}^{\infty}A_{l}r^{l}P_{l}\left(\cos\theta\right)\,.\label{eq:ansatz_Phi_ND}
\end{equation}
By using that $1/|\mathbf{r}-\mathbf{b}|=\sum_{l=0}^{\infty}\left(r_{<}^{l}/r_{>}^{l+1}\right)P_{l}\left(\cos\theta\right)$ [S1,\,S2] with $r_\gtrless$ being the greater (smaller) of $|\mathbf{r}|$ and $|\mathbf{b}|$, one can derive the potential in the electrolyte without discontinuity, i.e. without nanodiamond, to be
\begin{equation}
    \tilde{\Phi}^{e}\left(r,\theta\right)=\frac{q}{4\pi\epsilon_{0}\epsilon_{e}}\sum_{l=0}^{\infty}\frac{r_{<}^{l}}{r_{>}^{l+1}}P_{l}\left(\cos\theta\right)\,.
\end{equation}
The general solution would then be given as a superposition of this expression with Eq.\,(\ref{eq:Phi_ansatz}), i.e.  $\Phi^e = \tilde{\Phi}^{e} + \Phi$, which reads
\begin{equation}
    \Phi^{e}\left(r,\theta\right)=\sum_{l=0}^{\infty}\left(C_{l}\frac{1}{r^{l+1}}+\frac{q}{4\pi\epsilon_{0}\epsilon_{e}}\frac{r_{<}^{l}}{r_{>}^{l+1}}\right)P_{l}\left(\cos\theta\right)\,,\label{eq:ansatz_Phi_e}
\end{equation}
where it was used that in this case $A_l=0$ to ensure a vanishing potential at infinite distances to the origin, i.e. $\Phi^e \rightarrow 0$ for $r\rightarrow\infty$. The constants $A_l$ and $C_l$, which enter into, respectively, Eq.\,(\ref{eq:ansatz_Phi_ND}) and Eq.\,(\ref{eq:ansatz_Phi_e}), can be determined by requiring continuity at the interface between electrolyte and nanodiamond,
\begin{align}
    \left(\epsilon_e \mathbf{E}^e - \epsilon_{\mathrm{ND}} \mathbf{E}^{\mathrm{ND}}\right) \cdot \mathbf{n}_{\mathrm{ND}} &= 0 \\
    \left(\mathbf{E}^e - \mathbf{E}^{\mathrm{ND}}\right) \times \mathbf{n}_{\mathrm{ND}}\,,
\end{align}
where $\mathbf{n}_{\mathrm{ND}}=\mathbf{r}/r$ is the unit vector normal to the surface of the nanodiamond. These boundary conditions are satisfied, if
\begin{align}
    A_{l}&=\frac{q}{4\pi\epsilon_{0}\epsilon_{e}}\frac{1}{b^{l+1}}\frac{\epsilon_{e}\left(2l+1\right)}{\epsilon_{\mathrm{ND}}l+\epsilon_{e}\left(l+1\right)}\\
    C_{l}&=\frac{q}{4\pi\epsilon_{0}\epsilon_{e}}\frac{lr_{\mathrm{ND}}^{2l+1}}{b^{l+1}}\frac{\epsilon_{e}-\epsilon_{\mathrm{ND}}}{\epsilon_{\mathrm{ND}}l+\epsilon_{e}\left(l+1\right)}\,.
\end{align}
The electrostatic potential inside the nanodiamond therefore is
\begin{equation}
    \Phi^{\mathrm{ND}}\left(r,\theta\right)=\frac{q}{4\pi\epsilon_{0}\epsilon_{e}}\sum_{l=0}^{\infty}\frac{1}{b^{l+1}}\frac{\epsilon_{e}\left(2l+1\right)}{\epsilon_{\mathrm{ND}}l+\epsilon_{e}\left(l+1\right)}r^{l}P_{l}\left(\cos\theta\right) \label{eq:Phi_ND}
\end{equation}
and the electric field at the center, i.e. for $r=0$, can be calculated as
\begin{equation}
    \mathbf{E}\left(r=0,\theta\right)=\frac{q}{4\pi\epsilon_{0}}\frac{3}{2\epsilon_{e}+\epsilon_{\mathrm{ND}}}\frac{\mathbf{b}}{b^{3}}\,,\label{eq:E_center_ND}
\end{equation}
if it is used that in cartesian coordinates one has $e_{z}=\cos\theta e_{r}-\sin\theta e_{\theta}$ with $e_z$ the azimuthally symmetric unit vector and $e_r$ and $e_\theta$ the radial and altitudinal unit vectors. 

\subsection{Electric field variance}\label{App:SigmaE}
The probability of an ion to be located at $\mathbf{b}$ witin a sphere of radius $R$ around the nanodiamond is
\begin{equation}
    p\left(\mathbf{b}\right)=\begin{cases}
\frac{3}{4\pi}\frac{1}{R^{3}-r_{\mathrm{ND}}^{3}}, & r_{\mathrm{ND}}\leq b\leq R\\
0, & \text{otherwise.}
\end{cases}
\end{equation}
It can be easily verified that this distribution is normalized, i.e. $\int_{\mathbb{R}^3}\mathrm{d}^3\mathbf{b}\,p\left(\mathbf{b}\right)=1$. Direct calculation reveals $\langle E_z \rangle = 0$ and therefore
\begin{align}
    \sigma_{E_z,ion}^2 &= \langle E_z^2 \rangle \nonumber\\
    &= \frac{9q^{2}}{\left(4\pi\epsilon_{0}\right)^{2}}\frac{1}{\left(2\epsilon_{e}+\epsilon_{\mathrm{ND}}\right)^{2}}\frac{1}{R^{3}-r_{\mathrm{ND}}^{3}}\left(\frac{1}{r_{\mathrm{ND}}}-\frac{1}{R}\right)\,.
\end{align}
Under the assumption that the electric fields generated by the single ions are uncorrelated, the total fluctuations are given by multiplying the above expression with the number of ions inside the sphere. The standard deviation $\sigma_{E_{z}}^2 = cN_{A}V\sigma_{E_z,ion}^2$ of the electric field components with $N_A$ Avogadro's number, $c$ the molar ionic concentration and $V$ the volume in which the ions reside therefore is
\begin{equation}
    \sigma_{E_{z}} = \frac{|q|}{\epsilon_{0}\left(2\epsilon_{e}+\epsilon_{\mathrm{ND}}\right)}\sqrt{\frac{3N_{A}}{4\pi}}\sqrt{c\left(\frac{1}{r_{\mathrm{NV}}}-\frac{1}{R}\right)}\,.
\end{equation}
From this it can be seen that the expected electric field fluctuations increase with the molar concentration, i.e. $\sigma_{E_z}\propto \sqrt{c}$. 

\section{Hamiltonian in rotating frame}\label{App:H_RWA}
As derived by Doherty \textit{et al.} in Ref. [S3], the Hamiltonian of the NV-center in presence of axial magnetic fields $B_z$ and electric field components $E_i$ with $i=x,y,z$ and $\hbar=1$ is
\begin{align}
    \hat{\mathcal{H}}_{NV}&=\left(D+d_\parallel E_z\right)\hat{S}_z^2 + \gamma_e B_z \hat{S}_z \nonumber\\
    & + d_\perp \left[E_x\left(\hat{S}_y^2-\hat{S}_x^2\right)+E_y\left(\hat{S}_x\hat{S}_y+\hat{S}_y\hat{S}_x\right)\right]\,,\label{eq:H_NV_Doherty}
\end{align}
with $\gamma_e = 2.8\,\mathrm{MHz/G}$ the NV's gyromagnetic ratio [S4] and $d_\parallel = 0.35\,\mathrm{Hz\cdot cm/V}$ and $d_\perp = 17\,\mathrm{Hz\cdot cm/V}$ the axial and transverse dipole moments [S5]. By rewriting this Hamiltonian in terms of its frequency contributions $\beta_z = \gamma_e B_z$, $\xi_z = d_\parallel E_z$ and $\xi_\perp = d_\perp \sqrt{E_x^2+E_y^2}$ and by introducing the electric field polarization $\phi_E$, which defines the transverse electric field projections via $\xi_x = \xi_\perp \cos \phi_E$ and $\xi_y = \xi_\perp \sin \phi_E$, Eq.\,(\ref{eq:H_NV_Doherty}) can be rewritten as 
\begin{align}
     \hat{\mathcal{H}}_{NV} &= \left(D+\xi_z\right)\hat{S}_z^2 + \beta_z \hat{S}_z -\frac{\xi_\perp}{2}\left(e^{i\phi_E}\hat{S}_+^2 + h.c. \right)\,,
\end{align}
where $\hat{S}_\pm = \hat{S}_x \pm i\hat{S}_y$ are spin-1 ladder-operators and $h.c.$ means the hermitian conjugate. 

The NV-center can be driven by perpendicular (compared to the NV's symmetry axis) microwave magnetic fields of amplitude $\Omega = \gamma_e B_d$ and frequency $\omega_{d}$. To exert polarized drives onto the NV-center, two wires which are perpendicular to each other (see Fig.\,1(a) main text) are operated with a phase $\phi$ between each other. This drive can be described by an Hamiltonian [S6]
\begin{equation}
    \hat{\mathcal{H}}_d\left(t\right) = \Omega \left(\hat{S}_x\cos\left(\omega_{d}t\right) + \hat{S}_y\cos\left(\omega_{d}t + \phi\right)\right)\,.\label{eq:ansatz_Hd}
\end{equation}
Defining phase-factors $\epsilon_\pm\left(\phi\right) = \left(1-ie^{\mp i\phi}\right)/2$, similarly to Ref. [S6], allows to compactly account for different polarizations as $\epsilon_+=1$ only if $\phi=-\pi/2$ (i.e. right-circular polarization) and $\epsilon_-=1$ for left-circular polarized microwave fields ($\phi=+\pi/2$). By transforming $\hat{\mathcal{H}}_{NV} + \hat{\mathcal{H}}_d\left(t\right)$ into a frame oscillating with $\omega_{d}$ through the unitary $U=e^{i\omega_d S_z^2}$, one can derive the Hamiltonian under the rotating-wave approximation, which is
\begin{align}
    \hat{\mathcal{H}} &= \hat{\mathcal{H}}_0 + \hat{\mathcal{H}}_d \nonumber\\
    \hat{\mathcal{H}}_0 &= \left(\Delta + \xi_z\right)\hat{S}_z^2 + \beta_z \hat{S}_z - \frac{\xi_\perp}{2}\left(e^{i\phi_E}\hat{S}_+^2 + h.c.\right) \nonumber\\
    \hat{\mathcal{H}}_d &= \frac{\Omega}{\sqrt{2}}\left(\epsilon_-\ket{0}\bra{-1} + \epsilon_+ \ket{1}\bra{0} + h.c.\right)\,.\label{eq:H_rwa_appendix}
\end{align}

\subsection{Derivation of time-evolution operators}
To allow for the efficient calculation of pulse-sequences, time evolution operators of the free evolution $\hat{F}\left(\tau\right)$ and the drive $\hat{R}\left(T\right)$ will be derived in the following. 

\subsubsection{Free Evolution}
A possible set of eigenstates of $\hat{\mathcal{H}}_0$ is $\{ \ket{0},\ket{+},\ket{-}\}$ with
\begin{align}
    \ket{+}&=\cos\frac{\theta}{2}e^{i\phi_E/2}\ket{1}+\sin\frac{\theta}{2}e^{-i\phi_E/2}\ket{-1} \nonumber\\
    \ket{-}&=\sin\frac{\theta}{2}e^{i\phi_E/2}\ket{1}-\cos\frac{\theta}{2}e^{-i\phi_E/2}\ket{-1}\,,
\end{align}
where $\tan \theta = -\xi_\perp / \beta_z$,  with corresponding eigenenergies $\omega_0=0$ and $\omega_\pm = \Delta + \xi_z \pm \sqrt{\beta_z^2 + \xi_\perp^2}$. The time evolution operator of $\hat{\mathcal{H}}_{0}$ is $\hat{F}\left(\tau\right) = \sum_{i=\{0,\pm\}}e^{-i\omega_i \tau} \ket{i}\bra{i}$, where the sum is performed over all eigenstates of $\hat{\mathcal{H}}_0$. In the basis of $\{\ket{0}, \ket{\pm 1}\}$ this is
\begin{align}
    \hat{F}\left(\tau\right)&=\ket{0}\bra{0} + e^{-i\tau\left(\Delta+\xi_{z}\right)}\Bigl[\nonumber\\
    &i\frac{\xi_{\perp}}{x}\sin\left(\tau x\right)\left(e^{i\phi_{E}}\ket{1}\bra{-1} + h.c. \right) \nonumber\\
    &+\left(\cos\left(\tau x\right)-i\frac{\beta_{z}}{x}\sin\left(\tau x\right)\right)\ket{1}\bra{1}\nonumber\\
    &+\left(\cos\left(\tau x\right)+i\frac{\beta_{z}}{x}\sin\left(\tau x\right)\right)\ket{-1}\bra{-1}\Bigr]\,.\label{eq:Free_evolution}
\end{align}
Here the frequency of oscillation has been defined as $x=\sqrt{\beta_z^2+\xi_\perp^2}$.

\subsubsection{Microwave Drive}\label{App:Time-evolution_MW-drive}
To derive operators which describe the action of the microwave pulses, it will be assumed that these pulses exceed all other frequency scales in magnitude, i.e. $\Omega \gg \Delta,\beta_z,\xi_z,\xi_\perp$, such that $\hat{\mathcal{H}}\approx \frac{\Omega}{\sqrt{2}}\hat{\mathcal{\widetilde{H}}}_d$ with $\hat{\mathcal{\widetilde{H}}}_{d}=\left(\epsilon_{-}\ket{0}\bra{-1}+\epsilon_{+}\ket{1}\bra{0}+h.c.\right)$. By noting that $\hat{\mathcal{\widetilde{H}}}_{d}^3 = \hat{\mathcal{\widetilde{H}}}_{d}$, the time evolution
\begin{equation}
    \hat{R}\left(t\right) = e^{-it\hat{\mathcal{H}}_d} = \sum_{k=0}^{\infty}\frac{\left(\frac{-it\Omega}{\sqrt{2}}\right)^{n}}{n!}\left(\hat{\mathcal{\widetilde{H}}}_{d}\right)^{n}\,,
\end{equation}
can be calculated as
\begin{align}
    \hat{R}\left(t\right)&=\ket{1}\bra{1}\left(1-\left|\epsilon_{+}\right|^{2}\right)+\ket{-1}\bra{-1}\left(1-\left|\epsilon_{-}\right|^{2}\right)-\epsilon_{+}\epsilon_{-}\ket{1}\bra{-1}-\epsilon_{+}^{*}\epsilon_{-}^{*}\ket{-1}\bra{1}\nonumber\\
    &+\cos\left(\frac{t\Omega}{\sqrt{2}}\right)\left(\ket{0}\bra{0}+\left|\epsilon_{+}\right|^{2}\ket{1}\bra{1}+\left|\epsilon_{-}\right|^{2}\ket{-1}\bra{-1}+\epsilon_{+}\epsilon_{-}\ket{1}\bra{-1}+\epsilon_{+}^{*}\epsilon_{-}^{*}\ket{-1}\bra{1}\right)\nonumber\\
    &-i\sin\left(\frac{t\Omega}{\sqrt{2}}\right)\left(\epsilon_{-}\ket{0}\bra{-1}+\epsilon_{+}\ket{1}\bra{0}+h.c.\right)\,.
\end{align}
Depending on the polarization, one can induce Rabi oscillations between $\ket{0}$ and either $\ket{-1}$ for $\phi = \pi/2$ (denoted as $\hat{R}_+$) or $\ket{+1}$ ($\phi=-\pi/2$, $\hat{R}_-$),
\begin{align}
    \hat{R}\left(t\right)_\pm &= \ket{\mp 1}\bra{\mp 1} + \cos\left(\frac{\Omega t}{\sqrt{2}}\right) \Bigl(\ket{0}\bra{0} + \ket{\pm1}\bra{\pm1}\Bigr) \nonumber\\
    &-i \sin\left(\frac{\Omega t}{\sqrt{2}}\right) \Bigl(\ket{0}\bra{\pm1} + h.c. \Bigr)\,.\label{eq:R(t)_pm}
\end{align}
The system can be driven to both $\ket{\pm 1}$, if a linearly polarized drive is used,
\begin{align}
    R\left(t\right)_{0}&=\frac{1}{2}\left(\ket{1}\bra{1}+\ket{-1}\bra{-1}+i\ket{1}\bra{-1}-i\ket{-1}\bra{1}\right)\nonumber\\
    &+\cos\left(\frac{t\Omega}{\sqrt{2}}\right)\Bigl(\ket{0}\bra{0}\nonumber\\
    &+\frac{1}{2}\left(\ket{1}\bra{1}+\ket{-1}\bra{-1}-i\ket{1}\bra{-1}+i\ket{-1}\bra{1}\right)\Bigr)\nonumber\\
    &-\frac{1+i}{2}\sin\left(\frac{t\Omega}{\sqrt{2}}\right)\left(\ket{0}\bra{-1}+\ket{1}\bra{0}+h.c.\right)\,.
\end{align}
The last expression can similarly be compactly written by noting that $\left(1\pm i\right)/2 = e^{\pm i\pi/4}/\sqrt{2}$. These operators can then be used to describe the action of (polarized) $\pi$- and $\pi/2$-pulses onto the $\ket{m_s=0,\pm1}$-states of the NV-center.

\section{Fourier Transformation of FID-signal}\label{app:FT}
Some arbitrary signals $f$ and $\tilde{f}$ in time- and frequency-domain are connected to each other as
\begin{align}
    \tilde{f}\left(\omega\right) &= \mathrm{FT}\left[f\left(\tau\right)\right] = \int_{-\infty}^{+\infty}\mathrm{d}\tau\ f\left(\tau\right)e^{-i\omega \tau} \nonumber\\
    \mathrm{FT}^{-1}\left[\tilde{f}\left(\omega\right)\right] &= \frac{1}{2\pi}\int_{-\infty}^{+\infty}\mathrm{d}\omega\ \tilde{f}\left(\omega\right)e^{i\omega \tau}\,. \label{eq:Fourier_Transform}
\end{align}
To simplify the calculation of the Fourier transformed FID-signal, one can rewrite $\mathrm{FID_{\xi_\perp,\phi_E}}$ (Eq.\,(6) main text) as 
\begin{align}
    \mathrm{FID}_{\xi_z,\xi_\perp}\left(\tau\right) &= \frac{1}{4}\Bigl( \frac{3}{2} + \frac{1}{2}\cos\left(2\tau\xi_\perp\right) - \cos\left(\tau\left[\xi_\perp+\xi_z\right]\right)\nonumber\\ 
    &- \cos\left(\tau\left[\xi_\perp-\xi_z\right]\right) \Bigr)\,.
\end{align}
From Eq.\,(\ref{eq:Fourier_Transform}), one sees that $\mathrm{FT}\left[\cos\left(\tau x\right)\right] = \pi\left[\delta\left(x-\omega\right)+\delta\left(x+\omega\right)\right]$ and therefore
\begin{align}
    \widetilde{\mathrm{FID}}\left(\omega\right) &= \frac{\pi}{4}\Bigl( \frac{3}{2}\delta\left(\omega\right) + \frac{1}{2}\left[\delta\left(2\xi_\perp - \omega\right) + \delta\left(2\xi_\perp + \omega\right)\right] \nonumber\\ &-\left[\delta\left(\xi_\perp + \xi_z - \omega\right) + \delta\left(\xi_\perp + \xi_z + \omega\right)\right] \nonumber\\
    &-\left[\delta\left(\xi_\perp - \xi_z - \omega\right) + \delta\left(\xi_\perp - \xi_z + \omega\right)\right]\Bigr)\,.
\end{align}

\newpage

\section{Simulated pulse sequences for normally distributed electric fields}\label{App:Pulse_sequences_fluctuations}

\begin{figure}[h]
    \includegraphics[width=0.48\textwidth]{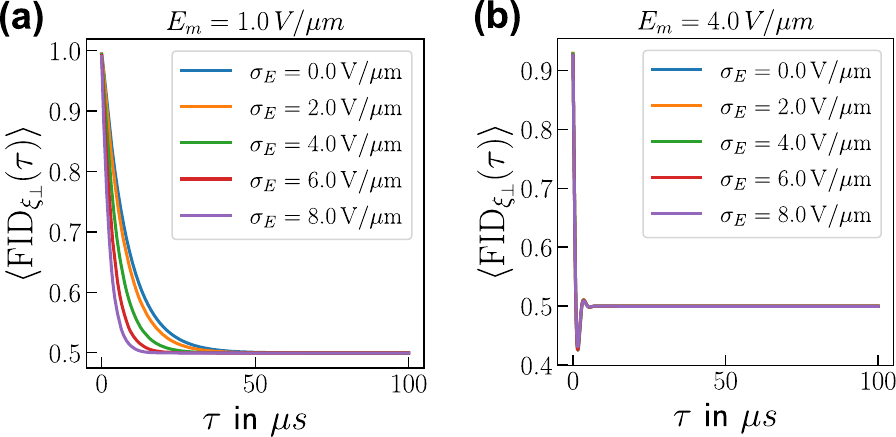}
    \caption{\label{fig:decay_fid_xi-perp}Simulated expected FID-values of $\mathrm{FID}_{\xi_\perp}$ (Eq.\,(5) main text), calculated from 500 individual FID-simulations with drive amplitude of $\Omega=10\,$MHz, intrinsic $T_{2,int.}^*$ and electric field components sampled from a normal distribution with mean $E_m$ and standard deviation $\sigma_E$. Dephasing is considered through a Lindblad-Operator $\sqrt{1/T_{2,int.}^*}S_z$. For both mean electric field values of (a) $1.0\,\mathrm{V/\mu m}$ and (b) $4.0\,\mathrm{V/\mu m}$, it is not possible to resolve $\xi_\perp$.}
\end{figure}

To understand how fluctuating electric fields alter the FID-signal, we numerically [S7,\,S8] simulated $\mathrm{FID}_{\xi_\perp}$ (Eq.\,(5) main text) for normally distributed electric fields. Hereby, at every timestep at which the time-evolution is calcuated, the electric field components are passed from a beforehand sampled normal distribution with mean $E_m$ and standard deviation $\sigma_E$. It can be seen from Fig.\,\ref{fig:decay_fid_xi-perp} that the average $\mathrm{FID}_{\xi_\perp}$ signal decays rapidly to its steady-state value of $1/2$, which is due to the short $T_2^*$ time of $1\,\mathrm{\mu s}$. For this reason it is proposed to use the Hahn-Echo pulse sequence for measurements of strongly fluctuating electric fields.

\begin{figure}[h]
    \centering
    \includegraphics[width=0.48\textwidth]{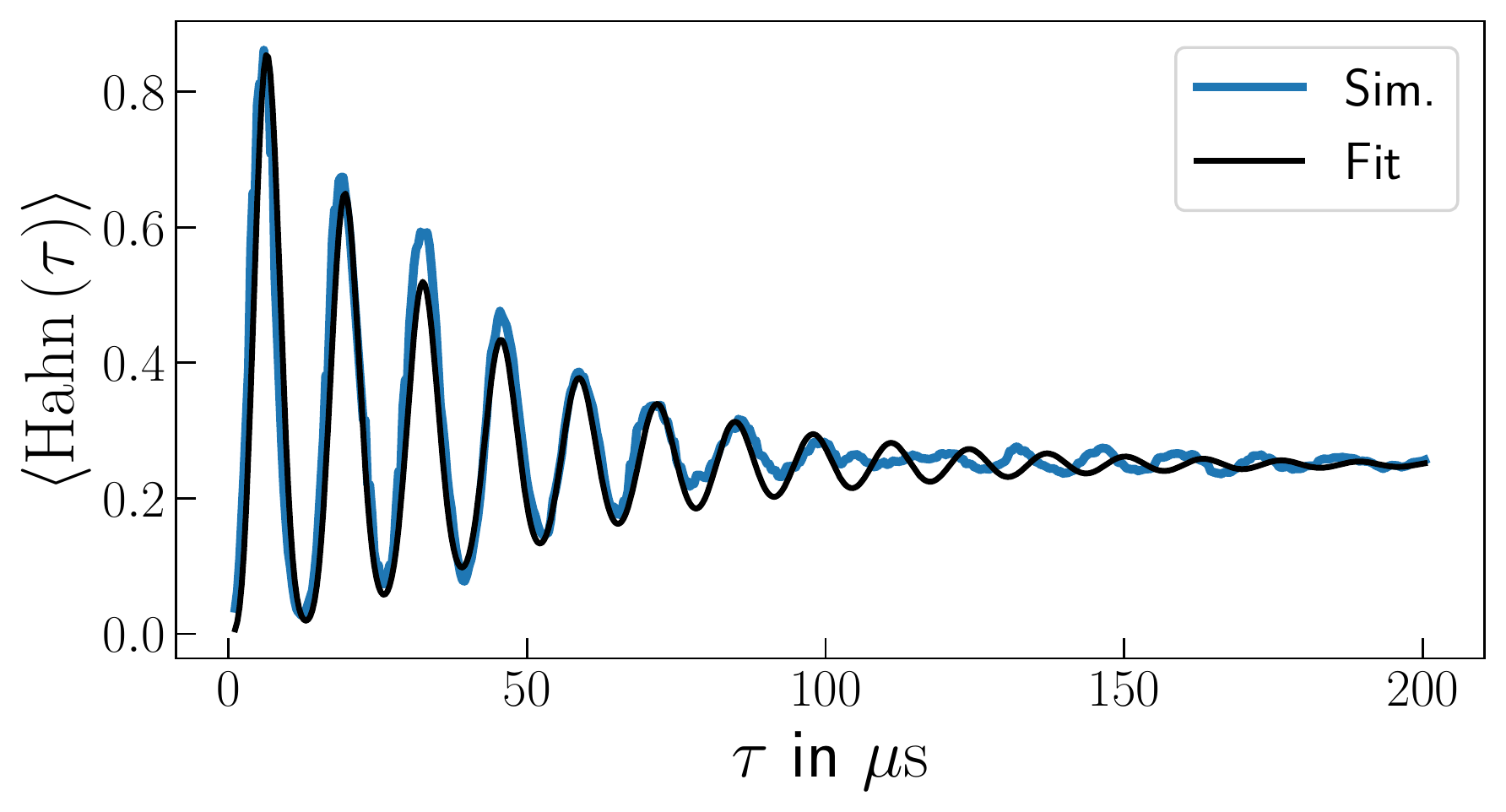}
    \caption{\label{fig:hahn-echo_example}Example of the average Hahn-echo signal, which was obtained numerically from 1000 individual simulations of the pulse sequence shown in Fig.\,3(a) (main text) with a mean electric field value of $E_m=1.0\,\mathrm{V/\mu m}$, standard deviation $\sigma_E=0.75\,\mathrm{V/\mu m}$, drive amplitude $\Omega=10\,\mathrm{MHz}$ and intrisic $T_{2,int.}=100\,\mathrm{\mu s}$ together with the fit following Eq.\,(10) (main text). The total $T_2$ value obtained from this fit is $T_2=\left(39.87 \pm 0.86\right)\,\mathrm{\mu s}$.}
    \label{fig:hahn-echo_signal}
\end{figure}

\begin{figure}[h]
    \centering
    \includegraphics[width=0.2\textwidth]{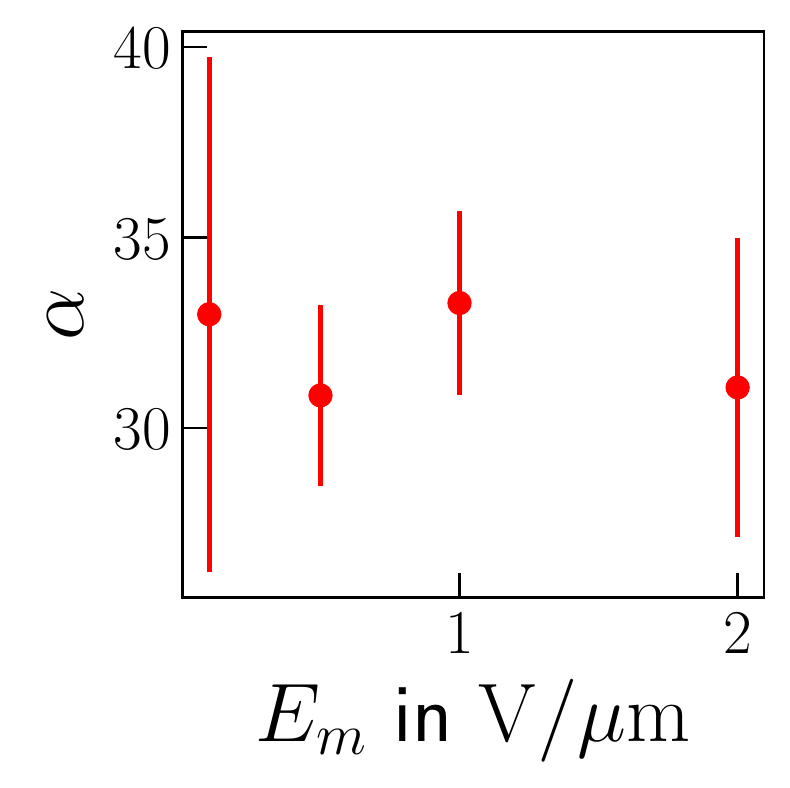}
    \caption{Fit parameter $\alpha$, obtained by numerically fitting Eq.\,(\ref{eq:T2_total}) and Eq.\,(\ref{eq:T2_E}) with $T_{2,int.}=100\,\mathrm{\mu s}$ to the data from Fig.\,3 (main text).}
    \label{fig:fit_parameter_hahn-echo}
\end{figure}

As described in the main text, the numerically obtained Hahn-echo trajectories (see Fig.\,\ref{fig:hahn-echo_example} for an example) are well fitted by $\left\langle \mathrm{Hahn}\left(\tau\right)\right\rangle = \frac{1}{4}\left[1-\cos\left(2\tau\xi_{\perp}\right)e^{-\tau/T_{2}}\right]^{2}$. Here both the intrinsic $T_{2,int.}=100\,\mathrm{\mu s}$ and $T_{2,E}$ due to fluctuating elecric fields contribute to the total $T_2$ via 
\begin{equation}
    \frac{1}{T_2} = \frac{1}{T_{2,int.}} + \frac{1}{T_{2,E}}\,. \label{eq:T2_total}
\end{equation}
The latter can be fitted in terms of $E_m$ and $\sigma_E$ via
\begin{equation}
    T_{2,E}=\alpha \frac{E_m}{\sigma_E^2}\,.\label{eq:T2_E}
\end{equation}
The values of the fit parameter $\alpha$ can be found in Fig.\,\ref{fig:fit_parameter_hahn-echo}.

\newpage
\renewcommand*\labelenumi{[S\theenumi]}

\begin{enumerate}
    \item R. Messina, ``Image charges in spherical geometry: Application to colloidal systems," \textit{The Journal of Chemical Physics,} vol. 117, no. 24, pp. 11062-11074, Dec. 2002.
    \item J. D. Jackson, ``Klassische Elektrodynamik," \textit{De Gruyter,} Dec. 2006.
    \item M. W. Doherty, F. Dolde, H. Fedder, F. Jelezko, J. Wrachtrup, N. B. Manson, and L. C. L. Hollenberg, ``Theory of the ground-state spin of the NV center in diamond," \textit{Physical Review B}, vol. 85, no. 20, p. 205203, May 2012.
    \item E. Abe and K. Sasaki, ``Tutorial: Magnetic resonance with nitrogen-vacancy centers in diamond - microwave engineering, materials science, and magnetometry," \textit{Journal of  Applied Physics}, vol. 123, no. 16, p. 161101, Apr. 2018.
    \item E. Van Oort and M. Glasbeek, ``Electric-field induced modulation of spin echoes of N-V centers in diamond," \textit{Chemical Physics Letters}, vol. 168, no. 6, pp. 529-532, May 1990.
    \item P. London, P. Balasubramanian, B. Naydenov, L. P. McGuiness, and F. Jelezko, ``Strong driving of a single spin using arbitrarily polarized fields," \textit{Physical Review A}, vol. 90, no. 1, p. 012302, July 2014.
    \item J. Johansson, P. Nation, and F. Nori, ``QuTiP: An open-source Python framework for the dynamics of open quantum systems," \textit{Computer Physics Communications}, vol. 183, no. 8, pp. 1760-1772, Aug. 2012.
    \item J. Johansson, P. Nation, and F. Nori, ``QuTiP 2: A Python framework for the dynamics of open quantum systems," \textit{Computer Physics Communications}, vol. 184, no. 4, pp. 1234-1240, Apr. 2013.
\end{enumerate}

\end{document}